\documentclass[11pt]{article} 

\usepackage[titlenumbered,ruled]{algorithm2e} 
\usepackage[margin=1.2in]{geometry}
\usepackage{natbib} 
\usepackage{graphicx} 
\usepackage{booktabs}
\usepackage{url} 
\usepackage{color} 
\usepackage{caption} 
\usepackage{subcaption} 
\usepackage{hyperref} 
\usepackage{authblk} 

\usepackage{amsmath,amsthm,amssymb} 
\usepackage{bbold} 

\DeclareMathOperator*{\argmin}{arg\,min}

\begin{document}

\title{{\Large Augmenting Adjusted Plus-Minus in Soccer with FIFA Ratings}}
\author[*]{Francesca Matano}
\author[*]{Lee F. Richardson}
\author[*]{Taylor Pospisil}
\author[*,**]{Collin Eubanks}
\author[*]{Jining Qin}
\affil[*]{Department of Statistics and Data Science, Carnegie Mellon University}
\affil[**]{Machine Learning Department, Carnegie Mellon University}
\date{\today}
\maketitle

\begin{abstract}
  In basketball and hockey, state-of-the-art player value statistics are often variants of Adjusted Plus-Minus (APM). But APM hasn't had the same impact in soccer, since soccer games are low scoring with a low number of substitutions. In soccer, perhaps the most comprehensive player value statistics come from video games, and in particular FIFA. FIFA ratings combine the subjective evaluations of over 9000 scouts, coaches, and season-ticket holders into ratings for over 18,000 players. This paper combines FIFA ratings and APM into a single metric, which we call Augmented APM. The key idea is recasting APM into a Bayesian framework, and incorporating  FIFA ratings into the prior distribution. We show that Augmented APM predicts better than both standard APM and a model using only FIFA ratings. We also show that Augmented APM decorrelates players that are highly collinear. 
\end{abstract}

\section{Introduction}
\label{sec:intro}
Decision making in sports typically requires player comparison. For example, teams decide which players to draft, who to offer a contract, etc. While not a panacea, one-number statistics of individual player value are useful decision making tools. With one-number statistics, teams can quickly rank free agents, evaluate trades, simulate future outcomes, and more. 

In basketball and hockey, the best one-number statistics are typically variants of Adjusted Plus-Minus (APM). APM is a regression technique that measures each players contribution to winning, while controlling for the quality of teammates and opponents. 

While {\it unadjusted} plus-minus has existed since the 1950's, {\it adjusted} plus minus was first used by Wayne Winston and Jeff Sagarin while consulting for the Dallas Mavericks (\cite{winval2003}). The first public APM calculation is due to \cite{rosenbaum2004measuring}, followed shortly after by \cite{ilardi2008adjusted}, who separated APM into offensive and defensive components. The next big advance was \cite{sill2010improved}, who replaced linear regression with ridge regression, and validated the improvement through out-of-sample predictions. APM is now mainstream in basketball; ESPN produces a ``Real Plus-Minus'' (RPM) statistic on their website (\cite{illardirpm2014}), and the RPM co-creator Jeremias Engelmann has written a book chapter on the subject (\cite{engelmann2017possession}). Recently, APM in basketball was extended to a ``win-probability'' framework (\cite{deshpande2016estimating}), which removes the effects of ``garbage time'' minutes. 

APM appeared in hockey shortly after basketball, with a series of papers by Brian Macdonald (\cite{macdonald2011regression,macdonald2011improved,macdonald2012expected,macdonald2012adjusted}). Since then, many variants have been proposed that adapt to the specifics of hockey. \cite{schuckers2013total} proposed the ``Total Hockey Rating'' (THoR), which models not only shots, but also events such as turnovers. \cite{gramacy2013estimating} uses logistic regression to model the probability of a goal scoring event, and \cite{thomas2013competing} uses hazard functions to model separate goal scoring processes for each team. 

\begin{figure}[ht]
  \centering
  \includegraphics[width = 14cm]{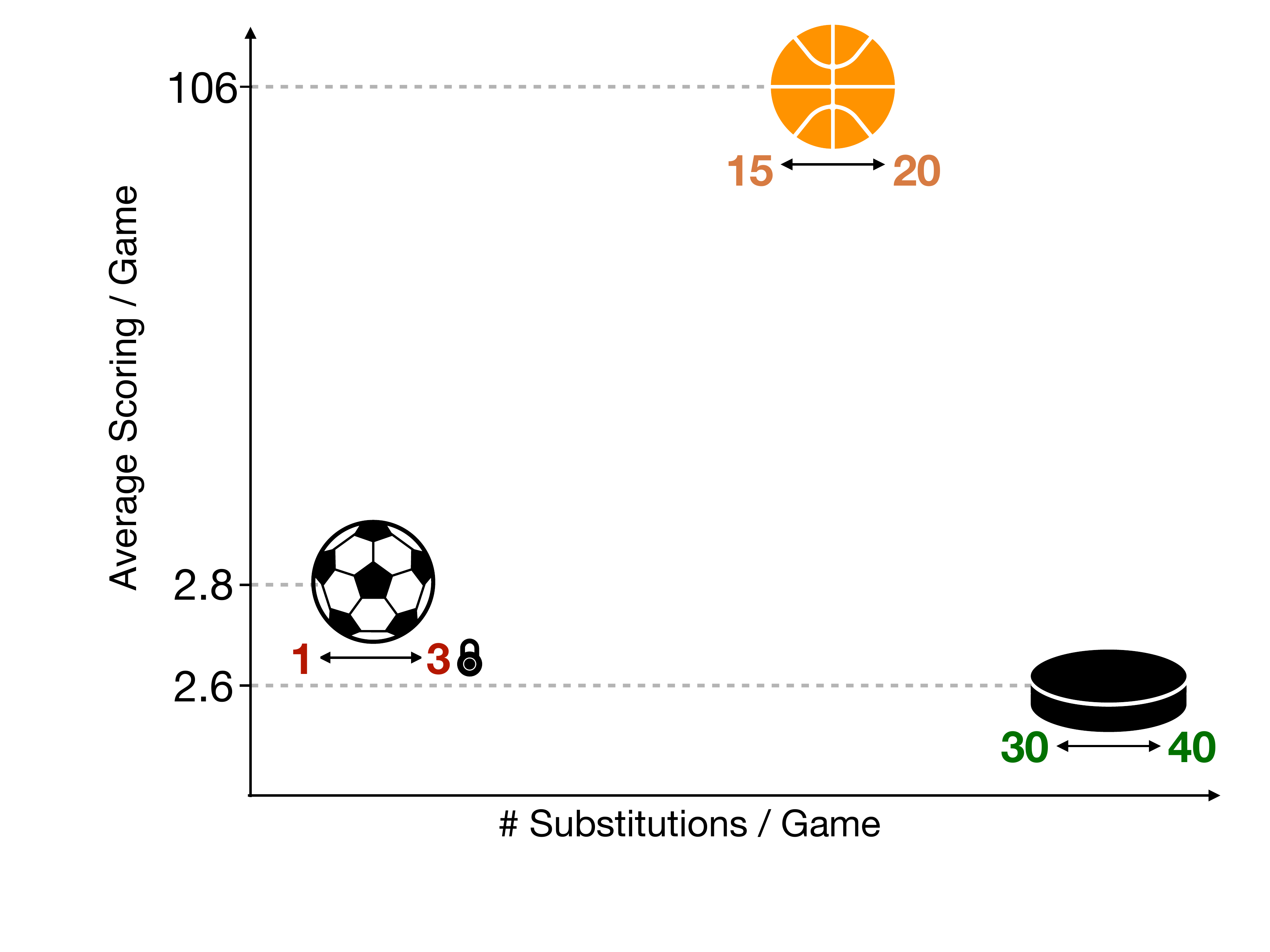}
  \caption{APM in basketball, hockey, and soccer, depends on the amount of scoring and substitutions. APM works best in basketball, which has by far the most scoring out of the three sports. Hockey has much less scoring than basketball, but has the most substitutions. Compared with hockey and basketball, soccer has both low scoring {\it and} low substitutions. This is the fundamental reason why APM is difficult in soccer.}
\label{fig:scoringsub}
\end{figure}

Compared with basketball and hockey, APM hasn't had a substantial impact in soccer. Since soccer has a low number of substitutions {\it and} a low number of scoring chances (see Figure \ref{fig:scoringsub}), standard APM has issues with collinearity and a sparse response variable. Two players who are almost always on the pitch together (consider two backs) will be indistinguishable to the APM model, since almost all of the segments which they play have been shared. Likewise, even for segments where one of these two backs is absent, we only observe a potentially small goal differential. 

That said, several authors have proposed versions of APM for soccer. \cite{ladder2009} first produced {\it unweighted} plus-minus statistics for the MLS in 2009. \cite{soccerstat2011} calculated {\it adjusted} plus-minus for the English Premier League in 2011, and cited \cite{rosenbaum2004measuring}, further demonstrating the cross-sport impact of APM. \cite{soccermetrics2014} gives a detailed calculation of APM in soccer, questions its utility, but concludes on an optimistic note, saying ``Adjusted plus/minus in football could become a valuable metric over time, but it will require a lot of care in its formulation, implementation, and interpretation.'' The first academic article on APM in soccer was \cite{kharrat2017plus}, who imports some of the best ideas from hockey and basketball into soccer. 

But, APM isn't the {\it only} way to compute one-number statistic. One popular alternative is video game ratings, for example from FIFA. FIFA provides ratings for over 18,000 players using the subjective evaluations of over 9,000 data-reviewers, who consist of coaches, scouts, and season ticket holders (\cite{fifaratings2016}). FIFA ratings are widely respected, and have recently been used to quantify the value of positions in soccer (\cite{pelechrinis2018positional}). In some sense, FIFA ratings are the epitome of the ``Intraocular'', or ``eye test''.

So, APM works well in basketball and hockey, but not soccer, and in soccer, FIFA ratings are respected one-number statistics. Putting it together, we propose Augmented APM, a Bayesian regression approach that combines FIFA ratings with APM.

The rest of this paper is organized as follows. Section \ref{sec:methods} details the Augmented APM model, Section \ref{sec:data} describes the data we use, Section \ref{sec:results} summarizes our results, and Section \ref{sec:discussion} concludes with a brief discussion.

\section{Methodology}
\label{sec:methods}
We extend the standard Adjusted Plus-Minus (APM) model, which solves the following ridge regression problem:

\begin{equation}
  \hat{\beta} = \argmin_{\beta} || y - X\beta ||_{2}^{2} + \lambda ||\beta||_{2}^{2}
\label{eq:ridge}
\end{equation}

We can recast this ridge regression problem as a maximum a posterior estimate from a Bayesian model

\begin{eqnarray}
\label{eq:apm}
  y | \beta &\sim& N(X\beta, \sigma^{2}) \nonumber \\
  \beta &\sim& N(0, \tau^{2})
\end{eqnarray}

for particular $\sigma$ and $\tau$ values. This recasting into a
Bayesian framework gives us several advantages:

\begin{itemize}
\item The tuning parameters become easier to fit and interpret. Under
  the optimization framework (Equation \ref{eq:ridge}), we use
  cross-validation to select an appropriate value of $\lambda$ that
  minimizes prediction error. Under a Bayesian interpretation, the
  similar parameter $\tau$ can be interpreted as the standard
  deviation of player abilities. This can either be selected
  intuitively (for the following results we set the values at $\tau=0.1$
  and $\sigma=1$) or given a hyperprior that specifies a wider 
  range of plausible values. 
\item We can obtain uncertainties for our estimates of $\beta$ by
  sampling from the posterior distribution. Many presentations of
  plus-minus scores neglect to publish measures of uncertainty in the
  estimates. This is particularly important in the context of
  comparing players as it's desirable to know whether differences in
  players reflect true differences in performance or simply a noisy
  estimation procedure. Likewise, we can express uncertainty in player 
  rankings by drawing from the joint posterior for betas.
\item We can easily extend the model to accommodate further
  information. We discuss two particular extensions of this model:
  including FIFA ratings and time-weighted segments. However,
  this does not exhaust the potential flexibility of these models, 
  which are often easy to fit using standard Bayesian 
  software (\cite{carpenter2017stan}).
\end{itemize}

We take advantage of the last point to incorporate subjective ratings into  the model. One problem with the traditional ridge penalty for players is that it assumes all players should be regularized towards zero: the effect of the ``average'' player. This is implicit in the prior which sets the prior mean for all players $\beta$ coefficients to be zero. But intuitively, priors should be different for more or less talented players: we would expect Lionel Messi to perform better than your typical player. Determining these prior values, however, could be difficult.

We can, however, use subjective assessments such as FIFA scores as an approximation for this prior. The Augmented APM model has the following multi-level form:

\begin{eqnarray}
\label{eq:ratingspm}
  y | \beta &\sim& N(X\beta, \sigma^{2}) \nonumber \\
  \beta | \alpha &\sim& N(\alpha \times \text{rating}, \tau^{2}) \nonumber \\
  \alpha &\sim& N(\mu_{\alpha}, \sigma_{\alpha}^{2}) 
\end{eqnarray}

Equation \ref{eq:ratingspm} is a hierarchical model where the
``prior'' for player ratings is no longer zeroed, and is instead
centered at a scaled value of their subjective ratings score. We shift
the ratings so that they have mean zero, which means that the $\beta$ values can
still be interpreted as an effect compared to the ``average'' player
(in the sense of having the average FIFA rating).

This model allows us to distinguish between players who are perfectly
collinear in the existing data set. While under a ridge regression
model both players would receive the same attribution of credit for
game results, the Augmented APM model gives the higher subjectively rated
player a greater share of the credit.

We also can introduce ``weights'' for the time length of a segment. Intuitively a longer segment will produce larger responses: scaling both the magnitude and variance of the differential $y$. To accommodate these effects we simply introduce the time of the segment into the model

\begin{eqnarray}
\label{eq:weighted_apm}
  y | \beta &\sim& N(tX\beta, t\sigma^{2}) \nonumber \\
  \beta &\sim& N(0, \tau^{2}).
\end{eqnarray}

This can be achieved in the standard APM optimization problem by estimating $yt^{-1}$ and introducing weights proportional to $\sqrt{t}$.

\section{Data}
\label{sec:data}
\begin{figure}
\centering
  \includegraphics[width = 14cm]{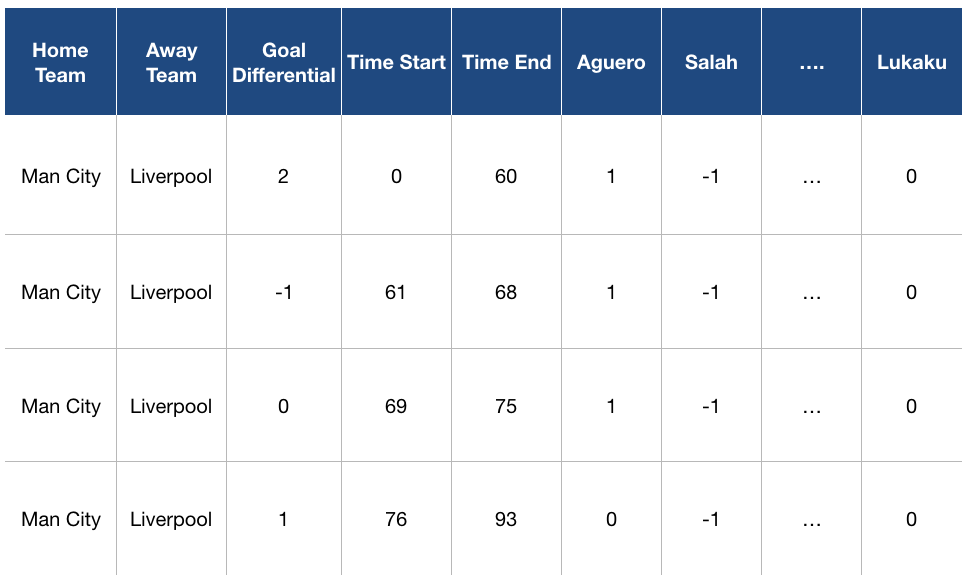}
  \caption{Play-by-play data, after pre-processing, for an example game between Manchester City and Liverpool. Each row corresponds to a game segment where no substitutions take place. The response variable is the goal differential in that game segment. Each player has a column, which is 1 when the player plays for the home team, -1 for the away team, and 0 if the player is not playing.}
\label{fig:pbp}
\end{figure}

We fit and evaluate our models on three English Premier League (EPL) seasons: 2015-16, 2016-17, and 2017-18. To fit these models, need two sources of data: play-by-play and FIFA ratings. Play-by-play data records all events (e.g. goals, substitutions) in a game. We convert this play-by-play data into a matrix where each row corresponds to a game segment without any substitutions. An example is shown in Figure \ref{fig:pbp}. 

We collect preseason FIFA ratings for each EPL season. The preseason FIFA ratings are typically released in late August of each year, on the website \url{sofifa.com}. While FIFA rates many player characteristics, such as speed and agility, this paper keeps things simple, and only uses each players overall rating. 

\section{Results}
\label{sec:results}
This section compares our Augmented APM model against the models listed in Table \ref{tab:models}. We use a criteria similar to \cite{sill2010improved}, in that we measure each models ability to predict out-of-sample game results. Game result predictions are simply the sum of predictions for each game segment (e.g. each row of Figure \ref{fig:pbp}). Our accuracy is the mean squared error (MSE) of the sum of predicted game segments with the actual game results. 

\begin{table}[ht]
\begin{tabular}{|l|l|}
\hline 
\textbf{Model} & \textbf{Description}  \\ \hline 
{\it Zero} & A naive model that predicts every game segment will \\ 
& have a $0$ goal differential. \\ \hline
{\it Intercept} & A naive model that predicts every game segment will \\ 
& have the average home advantage point differential  \\
& (learned on training data). \\ \hline 
{\it FIFA} & Uses the difference in teams overall FIFA ratings as predictors \\ \hline 
{\it APM} & Standard ridge regression APM (Equation \ref{eq:ridge}) \\ \hline 
{\it Augmented APM} & Equation \ref{eq:ratingspm} of Section \ref{sec:methods} \\ \hline 
\end{tabular}
\caption{The five models we compare in this paper.}
\label{tab:models}
\end{table}

\begin{figure}[h]
  \centering
  \includegraphics[width = 16cm]{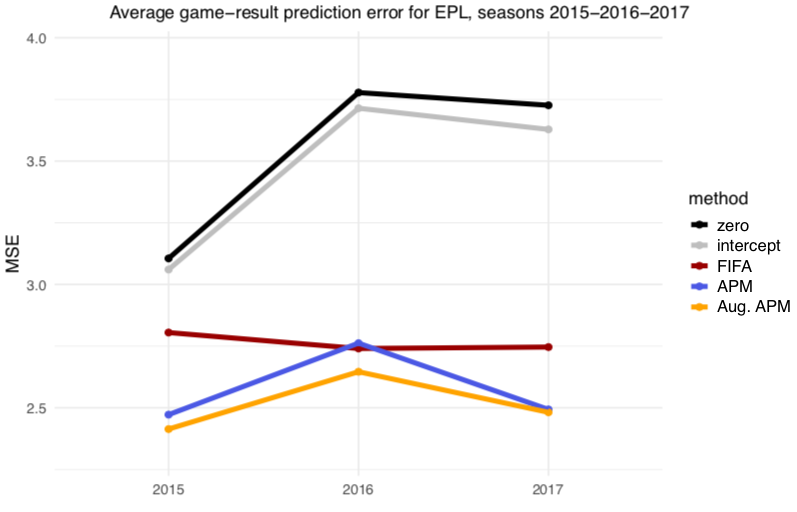}
  \caption{The MSE of our models in three EPL seasons using 10 fold cross-validation.}
  \label{fig:prediction}
\end{figure}

Figure \ref{fig:prediction} compares the models in each season using 10 fold cross-validation. Our Augmented APM model has the best predictive accuracy. The FIFA only model improves on the intercept model, which shows that FIFA ratings are a valuable predictor. We were surprised, however, that standard APM out-predicts the FIFA only model in 2015 and 2017, due to the limitations of APM in soccer discussed in the introduction. 

\begin{figure}[ht]
  \centering
  \includegraphics[width = 14cm]{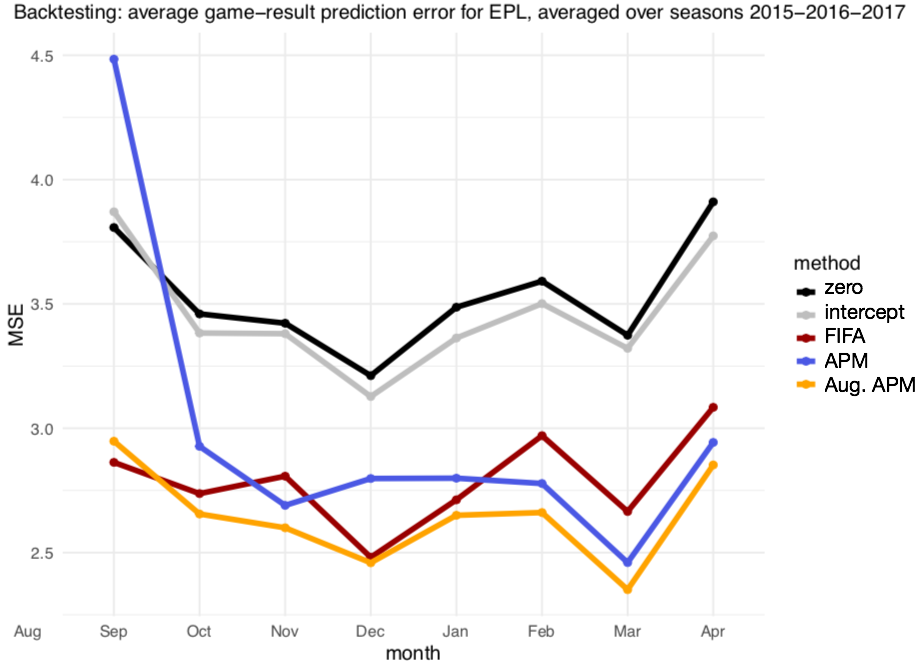}
  \caption{The MSE of each model over the course of the season. We train our models all data up until a month, then measure the prediction error over the next two months. The MSE for each month is averaged across the three seasons in our dataset.}
\label{fig:backtesting}
\end{figure}

We were also interesting in how our models performed at different times in the season. Intuitively, we expected that FIFA would predict better in the beginning of the season, and APM would predict better at the end of the season. To test this, we trained each model with all data up until a particular month, then predicted game results for the next two months. The results of this are shown in Figure \ref{fig:backtesting}, which demonstrates that FIFA starts the season as the best predictor, and both APM and Augmented APM out-predict FIFA by February. This suggests APM picks us useful information over the course of the season. 

\begin{figure}[ht]
  \centering
  \includegraphics[width = 15cm, scale=1]{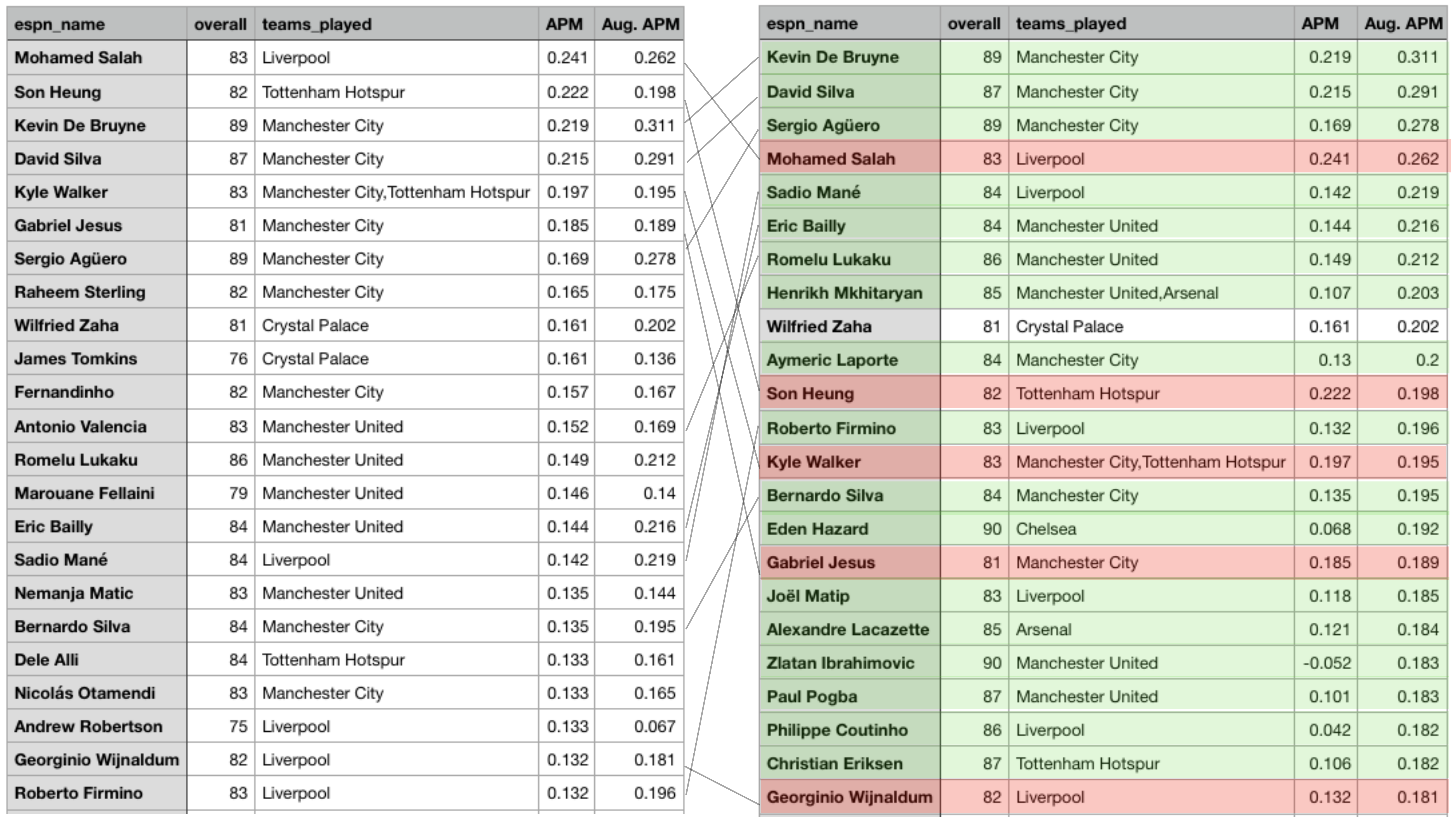}
  \caption{The top 15 players from the standard APM model (Left) compared with our Augmented APM model (Right). For the Augmented APM rankings, we color players with higher-than-APM ratings in green, and players with lower-than-APM ratings in red.}
\label{fig:intraocular}
\end{figure}

Adjusted Plus-Minus models are often validated by an ``Intraocular'' test, in that, it's better if the results make intuitive sense to soccer fans. Here, we expect the FIFA ratings to be very helpful. Figure \ref{fig:intraocular} compares the top 15 players from the APM and Augmented APM models in the 2017-18 season. As expected, Augmented APM gives higher value to players with higher FIFA ratings. We also see that Mohamed Salah, the EPL player of the year, ranks first in the APM model, and fourth in the Augmented APM model. This is a good sign for the intraocular test. 

Finally, an additional benefit of Augmented APM is that it ``de-correlates'' players who play most of their minutes together. To show this, Figure \ref{fig:intraocular} shows that players on Manchester City and Manchester United ``cluster'' together in the APM model. While some clustering still occurs in the Augmented APM ratings, it is less pronounced, since now players on the same team must {\it also} have similar FIFA ratings. 

\section{Discussion}
\label{sec:discussion}
This paper introduces Augmented APM, which incorporates FIFA ratings into the standard APM framework. The key idea was recasting APM into a Bayesian framework, then incorporating FIFA ratings into the prior distribution. We showed that Augmented APM has better out-of-sample prediction accuracy than models using only FIFA ratings, and standard APM. In addition, Augmented APM helps ``de-correlate'' players that play the majority of their minutes together. 

There are many directions for future work. One direction is simply adding more data and seasons into our estimates. This would allow us to rank players across leagues, especially if we used cross-league tournament data. Another direction we tried to include in this project is using ``Expected Goals'' instead of goal differential as the response variable. Unfortunately, our play-by-play data was too coarse, and our expected goals model did not boost our predictive accuracy. This is certainly an area where player tracking data would help. Finally, nothing conceptually prevents extending Augmented APM to other sports. 

We make both the software to generate the results, and the final results, available online. The software is in two separate {\tt R}-packages:

\begin{itemize}
  \item \href{https://github.com/tpospisi/PlusMinusModels}{{\tt PlusMinusModels}}. {\tt R}-package that fits the Augmented APM and APM models. 
  \item \href{https://github.com/fmatano/apm}{{\tt apm}}. {\tt R}-package that scrapes play-by-play data, FIFA ratings, and prepares the for modeling. 
\end{itemize}

\noindent The modeling results are available as sortable tables online at \url{www.intraocular.net/apm}.

\bibliographystyle{apa}
\bibliography{aapm-references}

\end{document}